\documentclass[traditabstract]{aa}
\usepackage{graphicx}
\usepackage[intlimits,sumlimits]{amsmath}
\usepackage{deluxetable}
\usepackage{times,epsfig} 
\usepackage{amssymb}
\usepackage{mathrsfs}  
\usepackage{color}
\usepackage{natbib}
\bibpunct{(}{)}{;}{a}{}{,} 
\usepackage{caption}
\usepackage{txfonts}
\usepackage{amstext}

\usepackage{natbib}

\usepackage{amsmath}
\usepackage[english]{babel}
\usepackage{longtable}
\usepackage{url}

\usepackage{hyperref}

\title{\textit{The first 48}: Discovery and progenitor constraints on the Type~Ia supernova 2013gy} 

\author{S. Holmbo\inst{1} 
\and M.~D. Stritzinger\inst{1,2}
\and B.~J. Shappee\inst{3}
\and M.~A. Tucker\inst{3}
\and W. Zheng\inst{4}
\and C. Ashall\inst{5}
\and M.~M. Phillips\inst{6}
\and C. Contreras\inst{6}
\and A. V. Filippenko\inst{4,7}
\and P. Hoeflich\inst{5}
\and M. Huber\inst{3}
\and X. F. Wang\inst{8}
\and J.-J. Zhang\inst{9,10}
\and J. Anais\inst{6}
\and E. Baron\inst{11,12}
\and C. R. Burns\inst{13}
\and A. Campillay\inst{6}
\and S. Castell\'on\inst{6}
\and C. Corco\inst{14}
\and E.~Y. Hsiao\inst{5}
\and K. Krisciunas\inst{15}
\and N. Morrell\inst{6}
\and M. T. B. Nielsen\inst{1}
\and S. E. Persson\inst{13}
\and A. Piro\inst{13}
\and F. Taddia\inst{16}
\and L. Tomasella\inst{17}
\and T.-M. Zhang\inst{19}
\and X.-L. Zhao\inst{8,20}
}

\institute{Department of Physics and Astronomy, Aarhus University, Ny Munkegade 120, DK-8000 Aarhus C, Denmark\\ (\email{simonholmbo@phys.au.dk, max@phys.au.dk})
\and
Visiting Astronomer, Institute for Astronomy, University of Hawai'i, 2680 Woodlawn Drive, Honolulu, HI 96822, USA 
\and
Institute for Astronomy, University of Hawai'i, 2680 Woodlawn Drive, Honolulu, HI 96822, USA
\and
Department of Astronomy, University of California, Berkeley, CA 94720-3411, USA
\and
Department of Physics, Florida State University, 77 Chieftain Way, Tallahassee, FL, 32306, USA
\and
Carnegie Observatories, Las Campanas Observatory, Casilla 601, La Serena, Chile
\and
Miller Senior Fellow, Miller Institute for Basic Research in Science, University of California, Berkeley, CA 94720
\and
Physics Department and Tsinghua Center for Astrophysics (THCA), Tsinghua University, Beijing 100084, China
\and
Yunnan Observatories, Chinese Academy of Sciences, Kunming 650216, China
\and
Key Laboratory for the Structure and Evolution of Celestial Objects, Chinese Academy of Sciences, Kunming 650216, China
\and
Homer L. Dodge Department of Physics and Astronomy, University of Oklahoma, 440 W. Brooks, Rm 100, Norman, OK 73019-2061, USA
\and
Visiting Astronomer, Hamburger Sternwarte, Gojenbergsweg 112, 21029 Hamburg, Germany
\and
Observatories of the Carnegie Institution for Science, 813 Santa Barbara St., Pasadena, CA 91101, USA
\and
Cerro Tololo Inter-American Observatory, National Optical Astronomy Observatory, Casilla 603, La Serena
\and
George P. and Cynthia Woods Mitchell Institute for Fundamental Physics and Astronomy, Department of Physics and Astronomy, Texas A\&M University, College Station, TX 77843, USA 
\and
The Oskar Klein Centre, Department of Astronomy, Stockholm University, AlbaNova, 10691 Stockholm, Sweden
\and
INAF, Osservatorio Astronomico di Padova, I-35122 Padova, Italy
\and
Yunnan Observatories, Chinese Academy of Sciences, Kunming 650216, China
\and
National Astronomical Observatories of China, Beijing, 100012, China
\and
School of Science, Tianjin University of Technology, Tianjin, 300384, China
}

\date{Received  05 September 2018 / Accepted XX XXXX 2018}

\abstract{We present an early-phase $g$-band light curve and visual-wavelength spectra of the normal Type~Ia supernova (SN)~2013gy.  
The light curve is constructed by determining the appropriate S-corrections to transform KAIT natural-system $B$- and $V$-band photometry and Carnegie Supernova Project natural-system $g$-band photometry to the Pan-STARRS1  $g$-band natural photometric system. 
A  Markov Chain Monte Carlo calculation provides a best-fit single power-law function to the first ten epochs of photometry described by an exponent  of $2.16^{+0.06}_{-0.06}$ and a  time of first light of  MJD 56629.4$^{+0.1}_{-0.1}$,  which is $1.93^{+0.12}_{-0.13}$ days (i.e., $<48$~hr) before the discovery date (2013 December 4.84 UT) and $-19.10^{+0.12}_{-0.13}$ days before the time of $B$-band maximum (MJD 56648.5$\pm0.1$).  The  estimate of the time of first light  is  consistent with the explosion time inferred from the  evolution of the \ion{Si}{ii} $\lambda$6355 Doppler velocity. Furthermore, discovery photometry and previous nondetection limits enable us to constrain the  companion radius down to $R_c \leq 4\,R_{\sun}$. 
In addition to our early-time constraints, we use a deep +235 day nebular-phase spectrum from Magellan/IMACS to place a stripped H-mass  limit  of  $< 0.018\,M_{\sun}$.  Combined, these limits effectively rule out  H-rich nondegenerate companions.
}

\authorrunning{Holmbo et al.}
\titlerunning{Early light curve of SN~2013gy}

\begin{document}

\maketitle

\begin{figure*}[ht]
\setcounter{figure}{2}
  \centering
  $\begin{array}{cc}
  \includegraphics[width=0.5\textwidth]{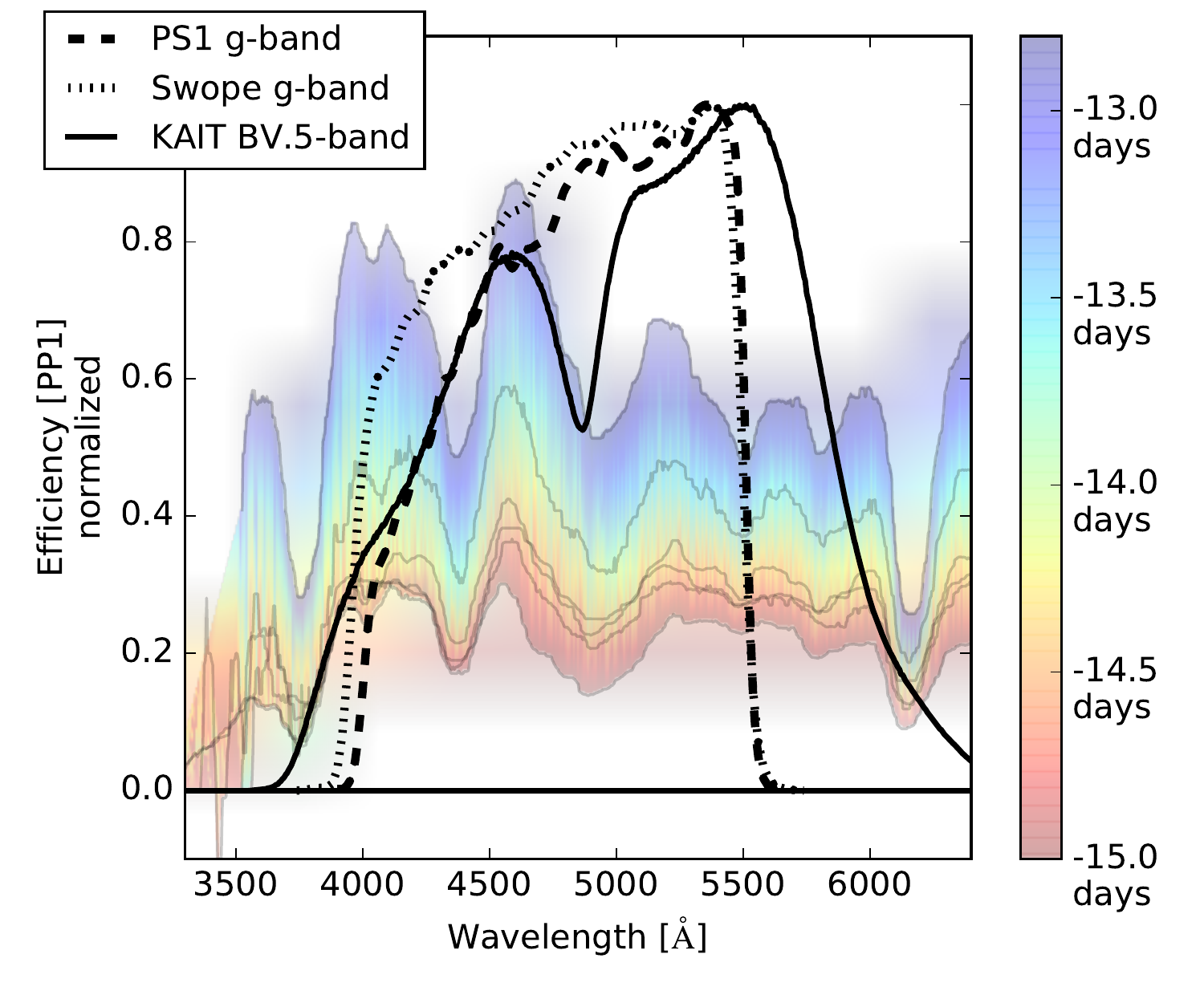}&
   \includegraphics[width=0.5\textwidth]{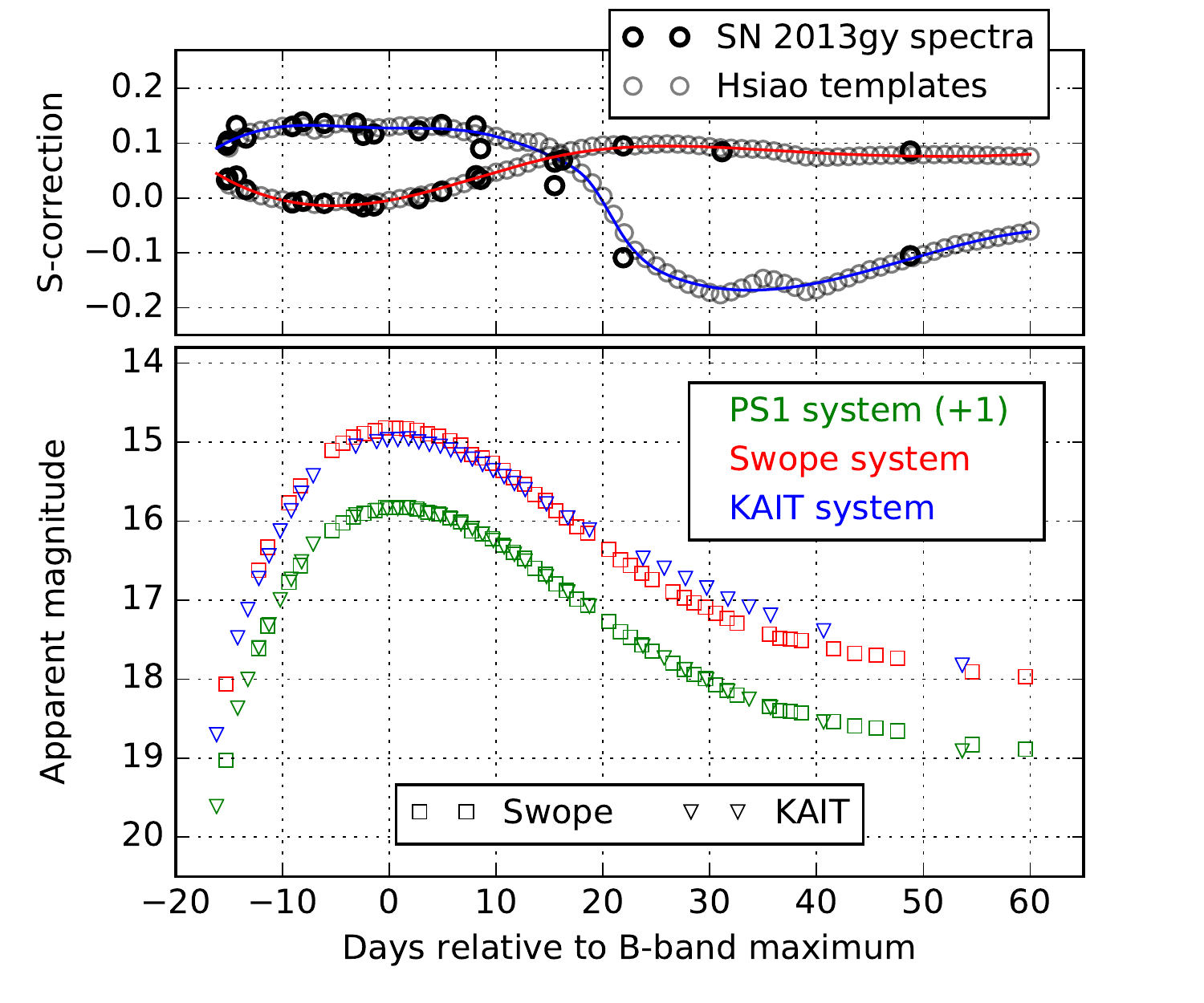}
   \end{array}$
 \caption{System response functions and S-correction results. \textit{(Left)} Comparison of the normalized $g_{\rm PS1}$ (dashed line), the $g_{\rm CSP}$ (dotted line), and $BV.5$ (solid line) system response functions. Also plotted are 7 visual-wavelength spectra of SN~2013gy ranging between $-16.1$~d to $-$13.4~d as indicated by the color bar.  
 \textit{(Right, top panel)}  S-corrections as a function of phase computing using observed (dark circles) spectra and \citet{hsiao07} template  spectra (light circles). The S-corrections enable us to accurately transform   $g_{\rm CSP}$-band  and $BV.5$-band natural-system  photometry to the $g_{\rm PS1}$  natural system.  
  \textit{(Right, bottom panel)} $g_{\rm CSP}$- (red squares) and $BV.5$-band (blue triangles) natural-system photometry, and the S-corrected versions (green points; offset by +1~mag) transformed to the $g_{\rm PS1}$ natural system.} 
  \label{fig:bandpasses}
\end{figure*}

\section{Introduction}\label{sec:introduction}
 
Today,  Type~Ia supernova (SN~Ia) cosmology is limited  by a subtle matrix of systematic errors. Although significant effort is being placed  to reduce known systematics in order to improve upon the accuracy of measuring their peak luminosities \citep[see][]{phillips18}, obtaining distances accurate to the percent level will likely require a better understanding of their progenitors  and explosion physics. 
Although SNe~Ia  likely originate from the thermonuclear disruption of a carbon-oxygen white dwarf \citep{hoyle60}, key aspects regarding the companion and explosion process remain poorly understood.
With the recent proliferation of very early discoveries and followup observations, SN~Ia
progenitor constraints can now be thoroughly tested. 
 Moreover, when combined with nebular spectroscopy  and X-ray limits, today's most constraining objects fail to match key predictions of the popular single-degenerate model \citep[see, e.g.,][]{shappee18a}.  

The current early-time SN~Ia sample  exhibits at least two different morphologies.  In one case, the early light curves evolve exponentially and are well fit with a single power-law function.
Objects exhibiting such a behaviour include SN~2011fe \citep{zhang16},  KSN~2011b \citep{olling15,shappee18b}, KSN~2012a \citep{olling15,shappee18b}, ASASSN-14lp (\citealt{shappee16}), and SN~2015F \citep{im15,cartier17}.
 On the other hand,  some SNe~Ia  exhibit a two-phase morphology.
In such cases the first phase consists of a linear increase in flux typically extending over 3--5 days past the time of first light ($t_{\rm first}$), followed by an abrupt exponential increase in flux as the light curve rises to maximum brightness.
Objects exhibiting a two-phase morphology  include SN~2012cg, \citep{marion16,shappee18a}, SN~2012fr \citep{contreras18}, SN~2013dy \citep{zheng13,pan15},  SN~2014J \citep{zheng14,siverd15,goobar15}, iPTF~16abc \citep{miller18}), and SN~2017cbv \citep{hosseinzadeh17}.
A two-component power-law function (or broken power-law function; \citealt{zheng17}) provides superior fits to the early light curves of these objects and more accurate estimates of both $t_{\rm first}$ and the rise time ($t_{\rm rise}$) to  maximum ($t_{\rm max}$).

Aside from four SNe~Ia located in Kepler fields \citep{olling15, brown18, cornect18}, today's sample of objects studied at early times consists of low-cadence coverage, and it may be possible that a two-component emission structure has been misinterpreted as a single power law.  For example, \citet{foley12} fit the early light curve of SN~2009ig with a single power law; however, when photometry computed from an open-filter discovery image is included, a two-component power law may indeed provide a superior fit \citep[see][]{contreras18}.
Taken together with the recent observations of MUSSES1604D (also known as SN~2016jhr; \citealt{jiang17}) a SN~2006bt-like event \citep{foley10, stritzinger11} that exhibited clear departures from a single/double power-law rise, the early-phase observational parameter space of SNe~Ia is far from fully explored.

\begin{figure*}[ht]
\setcounter{figure}{3}
  \centering
  $\begin{array}{cc}
  \includegraphics[width=0.5\textwidth]{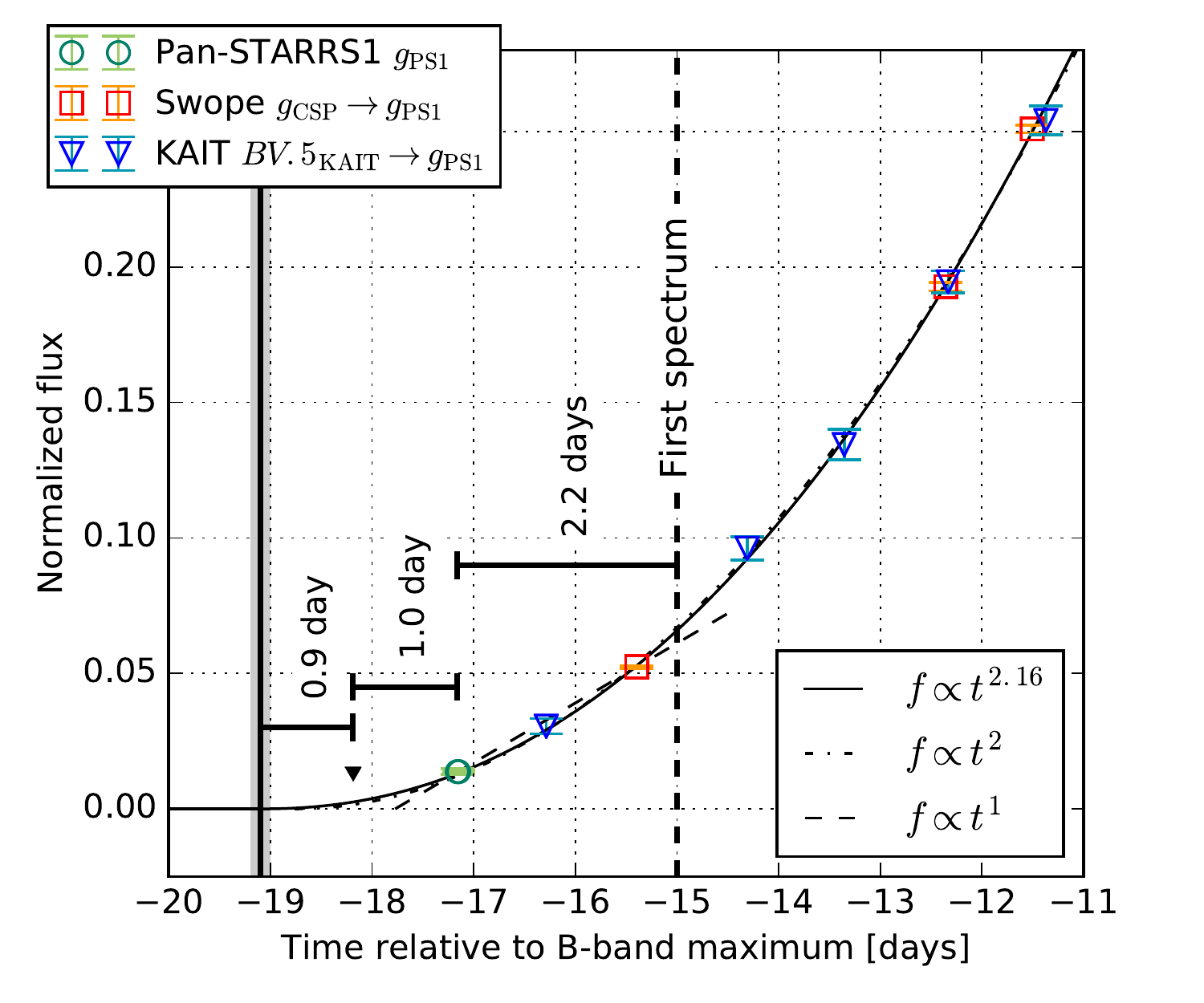}&
    \includegraphics[width=0.5\textwidth]{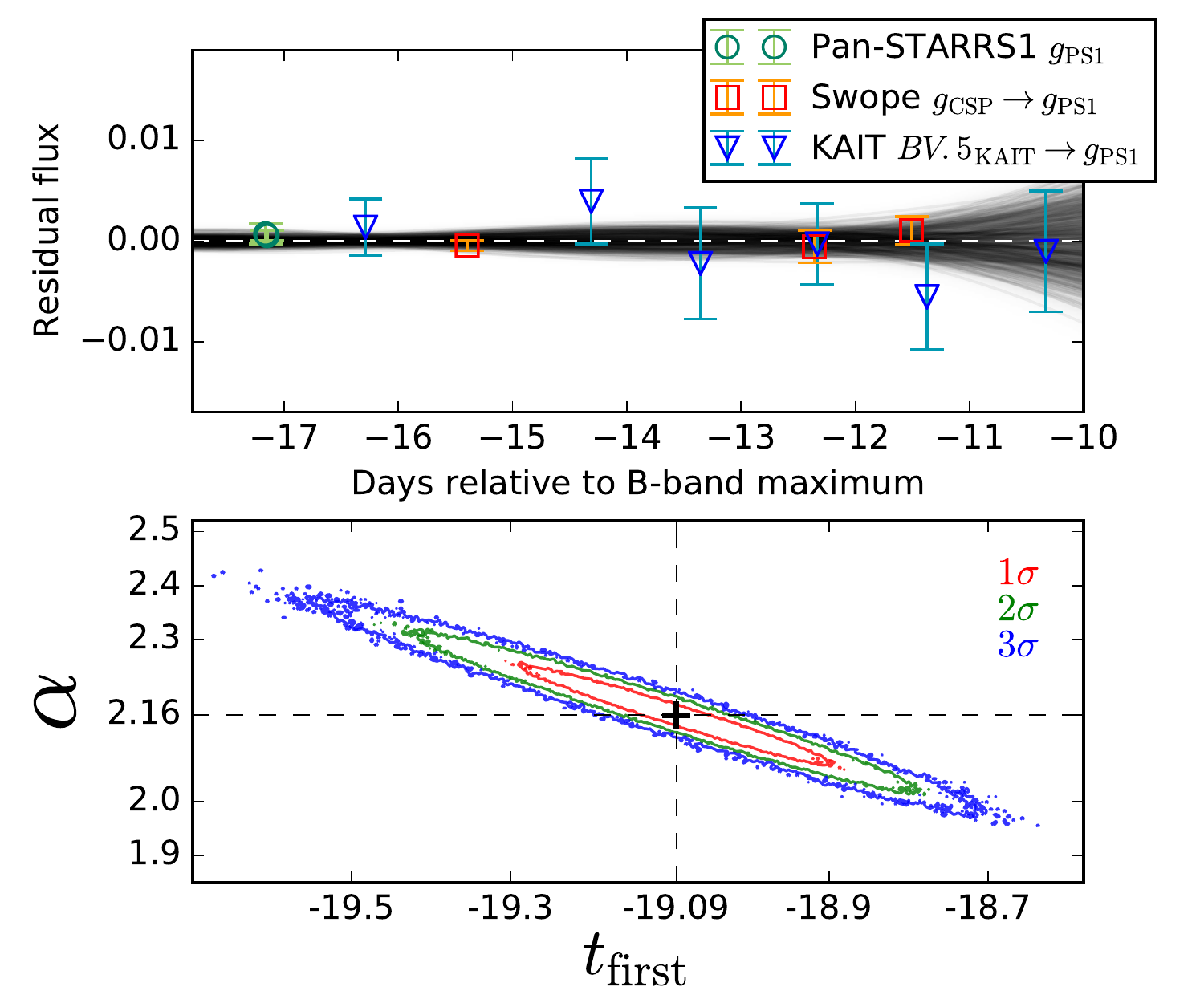}
    \end{array}$
  \caption{\textit{(Left)} The early-time $g_{\rm PS1}$-band light curve of SN~2013gy based on photometry measured from PS1, CSP-II, and KAIT images, plotted along with the best-fit, single power-law function characterized by an index $\alpha = 2.16$ (solid curve), the ``fireball" model ($\alpha = 2$, dash-dot curve), and a linear rise for the first $\sim 3$~d (dashed curve). The KAIT open-filter nondetection obtained 1~d prior to discovery is plotted as a black filled triangle. The solid vertical black line indicates $t_{\rm first}$ with its width corresponding to the 1$\sigma$ uncertainty (grey region), and the vertical dashed line corresponds to the epoch of the first spectrum. \textit{(Right, top)} Residuals between the best-fit model and the data (horizontal dashed white line); black lines correspond to models sampled from below (right, bottom panel), which contains $1\sigma$, $2\sigma$, and $3\sigma$ error ellipses indicating the uncertainties in the model fit parameters $\alpha$ and $t_{\rm first}$.}
  \label{fig:lightcurve}
\end{figure*}

Currently there are  $<$ 20  SNe~Ia in the literature discovered within $\sim$ 3 days of $t_{\rm first}$, and the addition of more  objects to the sample  is crucial to fully map out the early-time parameter space.
Here we present an early $g$-band light curve of the normal SN~2013gy (also known as PS1-13ejo) based on photometry calibrated to one well-understood photometric system.  Specifically,  Pan-STARRS1 (PS1; \citealt{tonry12,chambers16}) first $g$-band detection images are combined with follow-up imaging obtained by the Carnegie Supernova Project-II (CSP-II; \citealt{phillips18}) and the Lick Observatory Supernova Search (LOSS; \citealt{2001ASPC..246..121F}).
In addition to the light curve, a comprehensive set of visual-wavelength spectra is also presented. 

\citet{kim13} announced the discovery of SN~2013gy in open-filter images obtained with the 0.76-m Katzman Automatic Imaging Telescope (KAIT) at Lick Observatory on 2013 December 6.33 (UT dates are used throughout this paper). An open-filter image taken the previous day provides a nondetection limit of 19.3 mag. 
SN~2013gy was also recovered in  PS1 images obtained on 2013 December 5.34 with an apparent $g$-band (hereafter $g_{\rm PS1}$) magnitude of 19.48$\pm$0.09. The previous PS1 nondetection image was obtained 5.95 days earlier on 2013 November 29.39 with a limiting $g_{\rm PS1}$ magnitude of 20.54.
An optical spectrum obtained by \citet{tomasella13} on 2013 December 7.5 indicated the object was a young SN~Ia.  

With J2000 coordinates of $\alpha=03^\mathrm{h}42^\mathrm{m}16^\mathrm{s}.88$ and $\delta=-04^\circ43'18''52$, SN~2013gy was located 32$\arcsec$ North and 11$\arcsec$ East from the center of the SB(s)b host galaxy NGC~1418. 
A color image of NGC~1418 with the position of SN~2013gy indicated is provided in Fig.~\ref{fig:findingchart}. 
The redshift of NGC~1418 is $z = 0.014023$ \citep{catinella05} and its Tully Fisher (TF) distance ranges from 43.8 Mpc \citep{springob07}  to 88.8 Mpc \citep{theureau07}. Given the high dispersion of the TF distances, in the following we adopt the redshift distance, which after correcting for a Virgo, Great Attractor, and Shapley infall model and adopting $H_0 = 73$ km~s$^{-1}$~Mpc$^{-1}$ corresponds to  a distances of $55.9\pm3.9$ Mpc (i.e., $\mu = 33.75\pm0.15$ mag). This  is fully consistent with the distance derived from the comprehensive  set of optical and near-infrared (NIR) light curves obtained by the  CSP-II, which provides $\mu =  33.68\pm0.09$ mag (see Sect.~\ref{sec:LCparameters}).  

\section{Early-time observations of SN~2013gy}\label{sec:observations}
\subsection{Photometry and visual-wavelength spectroscopy}
\label{sec:photometry} 

Our early-phase $g_{\rm PS1}$-band light curve of SN~2013gy is constructed using data from three different facilities.  This includes two  epochs of $g_{\rm PS1}$ separated by $\sim20$~min,  52 epochs of $g$-band photometry obtained by the CSP-II (hereafter $g_{\rm CSP}$)  extending from $-$15.3~d\footnote{In keeping with tradition, temporal phases are given with respect to the time of $B$-band maximum brightness ($t_{B,{\rm max}}$) unless explicitly stated.} to $+$59.4~d, and  37 epochs of $B$- and $V$-band photometry from KAIT extending from $-$16.2~d to $+$53.6~d. To facilitate our analysis presented below, in the following the KAIT $B$- and $V$-band photometry is summed together (in flux space) and then multiplied by one half, creating what we refer to as $BV.5$-band photometry (see below). Reduced $g_{\rm PS1}$-band photometry was downloaded from the PS1 webpage\footnote{\url{https://star.pst.qub.ac.uk/ps1threepi/psdb/candidate/1034216880044318600/} \citep{smartt14}}, 
while CSP-II and KAIT images were reduced in the standard manner following the techniques described by \citet{krisicunas17} and \citet{zheng13}, respectively. Natural-system photometry of the supernova was computed for each set of science images relative to local sequences of stars. These  were  calibrated to the natural system of each setup using standard-star photometry converted to the natural systems  through the use of color terms obtained from multiple  observations of the standard fields \citep[see][for details]{contreras18}. The resulting $g_{\rm PS1}$-, $g_{\rm CSP}$-, and KAIT $BV.5$-band photometry is listed in Table~\ref{tab:photometry}, along with the S-corrections (see below) that transform their photometry to the $g_{\rm PS1}$ natural system.

\begin{figure*}[ht]
\setcounter{figure}{4}
  \centering
    \includegraphics[width=1\textwidth]{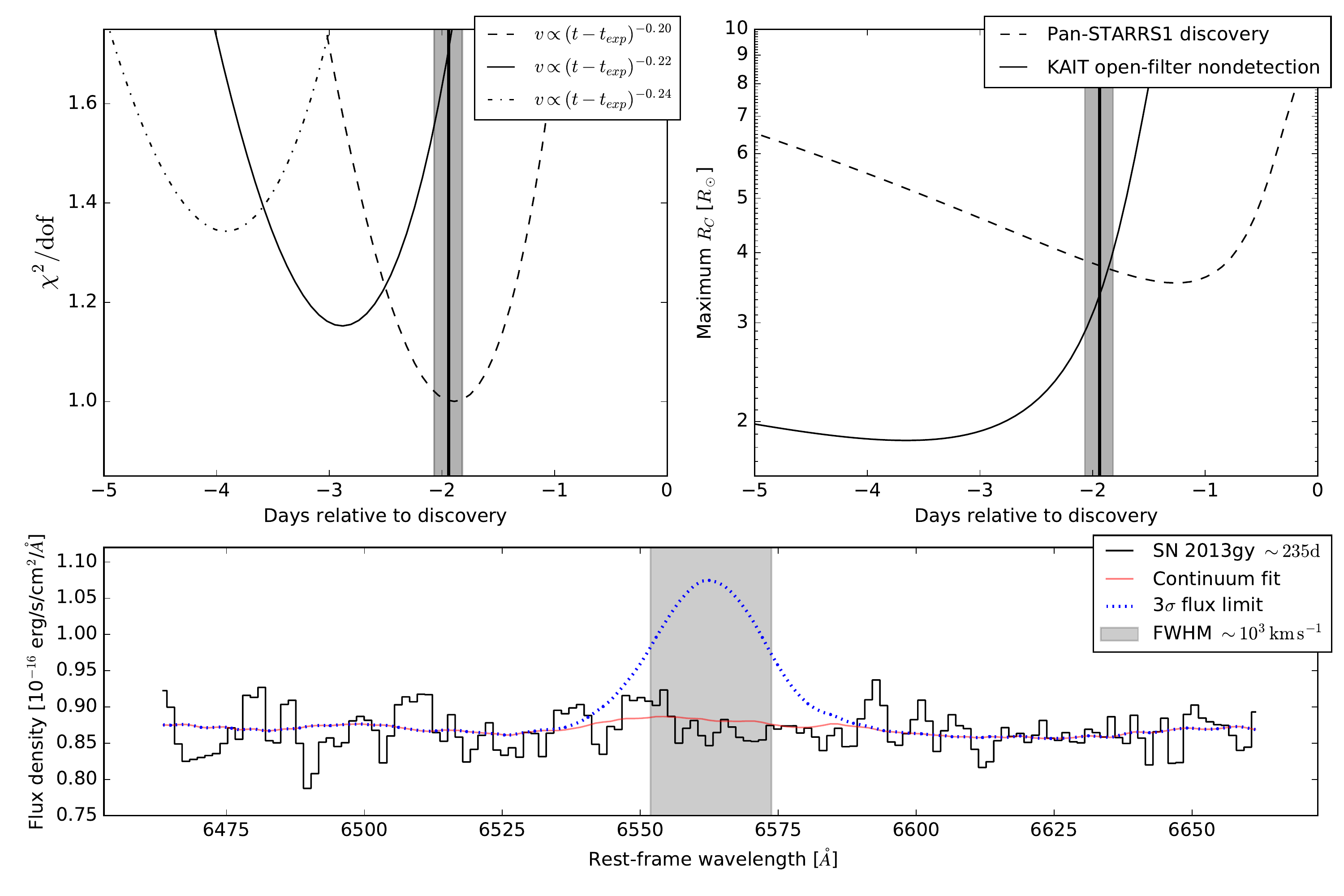}
  \caption{
 \textit{(Left)} Estimate of $t_{\rm explosion}$ based upon functional fits to the velocity evolution of $\ion{Si}{ii}$ $\lambda$6355. \textit{(Right)} Maximum allowed radius of a companion star as a function of explosion time as inferred from the  discovery $g_{\rm PS1}$ photometry and the previous KAIT nondetection limit plotted vs. days since discovery. Vertical black line in the left and right panels corresponds to the best-fit $t_{\rm first}$, accompanied by a shaded grey region corresponding to its 1$\sigma$ uncertainty. \textit{(Bottom)} Rest-frame-corrected nebular spectrum of SN~2013gy (black line), the continuum fit (red), and  our 3$\sigma$ H$\alpha$ flux limit (blue line).
The shaded area extends  over the expected location of H$\alpha$ $\pm$ 1000 km~s$^{-1}$.
}
  \label{fig:constraints}
\end{figure*}

Table~\ref{specjor} contains the journal of spectroscopic observations. These include  18 epochs of unpublished early optical spectra,  5 published spectra \citep{graham17}, and a  late-phase spectrum taken with the  Magellan Baade telescope.
 The spectra were reduced following standard procedures  \citep[see, e.g.,][]{hamuy06}, were color-matched to multiband photometry, and are plotted in the rest frame in  Fig.~\ref{fig:spectra}. 
 
\subsection{Light-curve parameters and reddening}
\label{sec:LCparameters}

Basic light-curve parameters were computed using the light-curve fitter \texttt{SNooPy} \citep{burns11}, and key results are listed in Table~\ref{results}.
\texttt{SNooPy} fits provide an estimate of the  time and magnitude of peak brightness, the light-curve decline-rate parameter $\Delta m_{15}(B)$, and the color-stretch parameter  $s_{BV}$ \citep{burns14}, as well as host-galaxy reddening. Using the \texttt{SNooPy} ``max model" $t_{B,{\rm max}}$ is found to have occurred on MJD = 56648.5$\pm0.1$ with $m_B$ $= 14.699\pm0.013$ mag, and $\Delta m_{15}(B) = 1.234\pm0.060$ mag.
Fitting the entire suite of CSP-II optical/NIR light curves\footnote{A detailed analysis of a comprehensive ultraviolet, optical, and NIR dataset of SN~2013gy will be presented in a forthcoming publication.} with  \texttt{SNooPy}'s ``EBV method2," we obtain an $s_{BV} = 0.892\pm0.05$  mag, a host-galaxy color excess of $E(B-V)_{\rm host} = 0.106\pm0.004_{\textrm{stat}}\pm0.060_{\textrm{sys}}$ mag, and a distance modulus $\mu =  33.68\pm0.09$ mag. Combining   $E(B-V)_{\rm host}$ with  the  Milky Way component of   $E(B-V)_{MW} = 0.049$ mag \citep{schlafly11,schlegel98}, we obtain $E(B-V)_{\rm tot} = 0.155\pm0.060$ mag, which is adopted throughout. 
Adjusting $m_{B}$ for reddening and  K-corrections, and assuming our adopted distance to the host, we find that SN~2013gy reached a peak absolute $B$-band magnitude $M_B = -19.32\pm0.16$ mag. 

\subsection{S-correction and a definitive $g_{\rm PS1}$ light curve}
\label{subsec:s-correction}

Here we describe how the $g_{\rm CSP}$- and $BV.5$-band photometry was transformed to the $g_{\rm PS1}$ natural system. We transform to the $g_{\rm PS1}$ natural system to avoid S-correcting the earliest observations where the S-corrections are more uncertain.
To compute accurate S-corrections requires system response functions and a spectrum, and/or a spectral template \citep{2002AJ....124.2100S}. 
The left panel of Fig.~\ref{fig:bandpasses} contains the $g_{\rm PS1}$-, $g_{\rm CSP}$-, and  KAIT $BV.5$-band response functions, along with early-phase optical spectra. 
The $BV.5$ response function is obtained by summing together the KAIT $B$- and $V$- response functions and multiplying the product by one half.
The motivation for doing this is that the S-corrections computed for the KAIT $B$ and $V$ bands were found to be very similar (on the order of $\pm0.6$ mag), but with opposite signs. So, by combining their photometry into a $BV.5$ system, the required S-corrections to transform the KAIT photometry to the $g_{\rm PS}$ system are minimized and found to lie between  $-0.1$ to $0.2$ mag out to $+$60~d.

To compute an S-correction at a specific epoch, an observed and/or template spectrum is first color-matched to CSP-II broad-band photometry. The S-correction that transforms $g_{\rm CSP}$-band natural-system photometry to the $g_{\rm PS1}$-band natural system is computed by taking the difference between synthetic photometry computed using the two different passbands and the spectrum --- that is, $m({g_{\rm PS1}}) = m(g_{\rm CSP}) - \Delta$S.  
Given that the spectral energy distribution of SNe~Ia is time dependent, S-corrections are computed for each epoch. Imaging was obtained using both the observed spectra and the \citeauthor{hsiao07} spectral template.
The S-corrections transforming our $g_{\rm CSP}$- and $BV.5$-band photometry to the $g_{\rm PS1}$ natural system are plotted in Fig.~\ref{fig:bandpasses} (right, top panel) and listed in Table~\ref{tab:photometry}.  The observed $g_{\rm CSP}$ and $BV.5$ photometry is also plotted (right, bottom panel)  with and without the S-corrections. 

\section{Results}
\label{sec:results}
\subsection{Early $g_{\rm PS1}$ light curve and constraints on $t_{\rm first}$}

Plotted in the left panel of Fig.~\ref{fig:lightcurve} is the early-time $g_{\rm PS1}$-band light curve of  SN~2013gy.  Overplotted on the light curve is the best-fit single power-law function (i.e., $f\left(t\right)=A\left(t-t_\mathrm{first}\right)^\alpha$), a single power-law function with an index of 2, and a linear function. The right, top panel shows the residuals between the best-fit model and the photometry. Finally, the right, bottom panel  of Fig.~\ref{fig:lightcurve} contains the estimated error in the fit parameters and their correlation. 
Fitting was performed using the Markov Chain Monte Carlo (MCMC) Python package \texttt{PyMC3}\footnote{\url{https://pymc-devs.github.io/pymc3}} \citep{2016ascl.soft10016S} and the No-U-Turn Sampler (NUTS; \citealt{2011arXiv1111.4246H}). 
When fitting we only considered data obtained prior to $-$10~d \citep{2006AJ....132.1707C,2011MNRAS.416.2607G}, and $\alpha$ was a free parameter with priors defined using the results  of \citet{2015MNRAS.446.3895F}. 
The best-fit power law corresponds to  $\alpha = 2.16^{+0.06}_{-0.06}$ and  $t_{\rm first} = 56629.4^{+0.1}_{-0.1}$ MJD. We note that from examination of the right, bottom panel of Fig.~\ref{fig:lightcurve}, a value of $\alpha = 2.0$ is found to be within 2$\sigma$ of the best-fit value. Assuming $\alpha = 2.0$ would imply a $t_{\rm first}$ value that is 7~hr later than our inferred best-fit value.
We also note that because the first 3 epochs
span more than 3 days and the uncertainty of the first KAIT measurement is relatively large, we cannot rule out a linear rise prior to $\sim15$~d before $t_{B,{\rm max}}$. The best fit to a linear rise implies a value of $t_{\rm first}$ that is 1.3~d later and only 14~hr before the PS1 images.
Comparison of the best-fit $t_{\rm rise}$ to the first PS1 detection indicates that SN~2013gy was discovered $1.93^{+0.12}_{-0.13}$~d past $t_{\rm first}$, and when compared to $t_{B,{\rm max}}$, one obtains $t_{\rm rise} = -19.10^{+0.12}_{-0.13}$ days.

\subsection{Spectroscopy and an estimate of $t_{\rm explosion}$}

Spectra of SN~2013gy taken around maximum light resemble those of normal SNe~Ia. Pseudo-equivalent-width (pEW) measurements of the \ion{Si}{ii} $\lambda$5972 and $\lambda$6355 features indicate that it is ``core normal'' (CN) according to \citet{branch06}. Moreover, the position of the absorption minimum of the $\ion{Si}{ii}$ $\lambda$6355 feature in the $-$0.8~d spectrum indicates a Doppler velocity at maximum absorption of $-v_{\rm abs} = 10,180\pm90$ km~s$^{-1}$, consistent with a normal object in the \citet{wang09} system.

\citet{piro13} suggested that an estimate of the explosion time ($t_{\rm explosion}$) can be obtained through the evolution of \ion{Si}{ii} $\lambda$6355. This is achieved by fitting the early-time velocity evolution with an  appropriate power law. Following \citeauthor{piro13}, we fit a $v \propto t^{\beta}$ power law with $\beta = -0.20$, $-0.22$, and $-0.24$. 
The resulting goodness-of-fit parameter $\chi^2$ per degree of freedom ($\chi^2$/dof) is plotted in the top-left panel of Fig.~\ref{fig:constraints} for each power-law index. A constant systematic error is added to get a best fit with $\chi^2$/dof $= 1$. We find that $\beta = -0.20$ provides a superior fit and implies $t_{\rm explosion} = 56629.45^{+0.28}_{-0.31}$ MJD.  
Comparison of  $t_{\rm first}$ and $t_{\rm explosion}$ reveals that SN~2013gy likely experienced minimal to no dark phase. 

\subsection{Constraints on interaction with a nondegenerate companion}

Companion radius ($R_c$) limits are computed following the methodology of \citet[][]{shappee18a}, which is based on comparing the observations to the analytical model predictions of \citet{kasen10}. 
Taking a conservative explosion time to be $\sim 2$~d before discovery and assuming a favourable viewing angle of $15^\circ$, we compute $R_c$ limits using both the discovery $g_{\rm PS1}$ photometry and the previous KAIT nondetection. Our limits on $R_c$ are plotted vs. days relative to the epoch of discovery in the top-right panel of Fig.~\ref{fig:constraints}. 
Combined limits indicate $R_c \lesssim 4\,R_{\sun}$ in the case of a favourable viewing angle.

\subsection{Limits on the presence of companion material}

Hydrodynamic simulations of ejecta-companion interactions indicate that $>0.15\,M_\odot$ of material could be removed from hydrogen-rich (H-rich) Roche-lobe overflow (RLOF) companions via stripping/momentum-transfer or ablation/heating and is expected to have a velocity dispersion of $v \approx 1000$ km~s$^{-1}$ (e.g., \citealp{marietta00, pan12, boehner17}). 
This unbound material should be visible during the nebular phase --- when the ejecta become optically thin --- and lead to a prominent emission signature. 

To place a limit on the H$\alpha$ emission we follow conventional techniques in the literature \citep[e.g.,][]{leonard07,shappee13}. 
Briefly, the late-phase ($+$235~d) spectrum of SN~2013gy  was flux calibrated  to match photometry computed from PS1 stars \citep{flewelling16} located in an $R$-band acquisition image.
Next, the continuum was  fitted in the vicinity of H$\alpha$ and subtracted from the spectrum, yielding no  statistical evidence for H$\alpha$ emission as demonstrated in the bottom panel of Fig.~\ref{fig:constraints}.
Following  \citet{leonard07}, we obtain a 3$\sigma$ statistical flux limit of $4.5\times 10^{-16}$ ergs~s$^{-1}$~cm$^{-2}$~\AA$^{-1}$. 
For our  adopted distance and using the models of  \citet{botyanszki18}, this corresponds to an H-mass limit of  $< 0.018\,M_\odot$.
This  limit is an order of magnitude lower than all estimates of unbound mass values for normal SNe~Ia, effectively ruling out H-rich RLOF companions. 
This leaves only a small parameter space of subdwarfs and He stars whose signatures could remain undetected in the spectrum. 

\section{Discussion}\label{sec:discussion}

 We have presented discovery and follow-up photometry  of the normal SN~Ia 2013gy. Photometry obtained from different facilities was carefully calibrated to the $g_{\rm PS1}$ system. The early rise of the light curve is well fit with a single power-law function, indicating that SN~2013gy was discovered within 48~hr of  $t_{\rm first}$.  
The corresponding $t_{\rm first}$  value is consistent with a $t_{\rm explosion}$ estimate based on the Doppler velocity evolution of the $\ion{Si}{ii}$ $\lambda$6355 feature. A short dark phase has been inferred from the study of several other supernovae including SN~2009ig, SN~2011fe, SN~2012cg \citep{piro14}, and iPTF~16abc \citep{miller18}. In contrast, the transitional  iPTF~13ebh may have experienced a $\sim4$~d dark phase \citep[see][]{hsiao15}.

Examination of the pEW values of the \ion{Si}{ii} $\lambda$5972 and $\lambda$6355 doublets shows that SN~2013gy  is core normal \citep{branch06,branch09}, while its \ion{Si}{ii} $\lambda$6355 Doppler velocity makes it a normal object according to the classification scheme of \citet{wang09}. Moreover, as discussed by \citet{stritzinger18}, the early intrinsic $(B-V)_{0}$ colors reveal it to be a red and rapidly evolving object, similar to SN~2011fe.  In summary, SN~2013gy displays all of the characteristics of a normal SN~Ia.

Making use of the early discovery of SN~2013gy, limits were placed on the radius of any companion star of $R \leq 4\,M_{sun}$. Combined with our stripped H-mass limit of $\lesssim 0.018\,M_{\sun}$ obtained from a nebular-phase Magellan spectrum, our analysis effectively rules out H-rich RLOF companions. This result highlights the  power to constrain existing SN~Ia progenitor models with the use of both early and late-phase observations.

\begin{acknowledgements}
We thank P. Massey and M. Kasliwal for obtaining a spectrum each with the Magellan telescopes.
S. Holmbo and M. D. Stritzinger are supported by a research grant (13261) from the VILLUM FONDEN. 
A. V. Filippenko's supernova group is grateful for financial assistance from USA National Science Foundation (NSF) grant AST-1211916, the TABASGO Foundation, the Christopher R. Redlich Fund, and the Miller Institute for Basic Research in Science (U. C. Berkeley). Research at Lick Observatory is partially supported by a generous gift from Google.
P. H\"{o}flich acknowledges NSF grants AST-1715133 and AST-1613472.
L. Tomasella is partially supported by the PRIN-INAF 2016 with the project ``Towards the SKA and CTA era: discovery, localisation, and physics of transient sources'' (P.I. M. Giroletti).

The CSP-II has been funded by the USA NSF under grants AST-0306969, AST-0607438, AST-1008343, AST-1613426, AST-1613455, and AST-1613472, and in part by  a Sapere Aude Level 2 grant  funded by the Danish Agency for Science and Technology and Innovation  (PI Stritzinger). J.-J. Zhang is supported by the National Natural Science Foundation of China (NSFC, grants 11403096 and 11773067), the Key Research Program of the CAS (grant KJZD-EW-M06), the Youth Innovation Promotion Association of the CAS (grant 2018081), and the CAS Light of West China Program. X. Wang  is supported by the National Natural Science Foundation of China (NSFC grants  11325313 and 11633002), and the National Program on Key Research and Development Project (grant 2016YFA0400803).

Based in part on observations made with the Nordic Optical Telescope, operated by the Nordic Optical Telescope Scientific Association at the Observatorio del Roque de los Muchachos, La Palma, Spain, of the Instituto de Astrofisica de Canarias. This paper includes data gathered with the 6.5~m Magellan telescopes located at Las Campanas Observatory, Chile.
We acknowledge the support of the staff of Lijiang 2.4~m and  the Xinglong 2.16~m telescopes. 
Partially based on observations made with the Copernico 1.8~m telescope (Asiago, Italy) operated by INAF Osservatorio Astronomico di Padova.

This paper made use of discovery images obtained by the Pan-STARRS1 Survey (PS1), which was made possible through contributions of the Institute for Astronomy, the University of Hawaii, the Pan-STARRS Project Office, the Max-Planck Society and its participating institutes, the Max Planck Institute for Astronomy, Heidelberg and the Max Planck Institute for Extraterrestrial Physics, Garching, The Johns Hopkins University, Durham University, the University of Edinburgh, Queen's University Belfast, the Harvard-Smithsonian Center for Astrophysics, the Las Cumbres Observatory Global Telescope Network Incorporated, the National Central University of Taiwan, the Space Telescope Science Institute, the National Aeronautics and Space Administration (NASA) under grant NNX08AR22G issued through the Planetary Science Division of the NASA Science Mission Directorate, the NSF under grant AST-1238877, the University of Maryland, and Eotvos Lorand University (ELTE).
\end{acknowledgements}

\clearpage
\begin{deluxetable}{crcrc}
\tablecolumns{3}
\tablewidth{0pt}
\tablecaption{Photometry of SN~2013gy and S-corrections.}\label{tab:photometry}
\tablehead{
\colhead{MJD} &
\colhead{Phase$^{a}$} &
\colhead{Magnitude} &
\colhead{S-correction} &
\colhead{$g_{PS1}$}}
\startdata
\multicolumn{5}{c}{$g_{PS1}$ band}\\
56631.34 & $-$17.16 & $19.483\pm0.079$ & $\cdots$ & $19.483\pm0.079$ \\
56631.35 & $-$17.15 & $19.482\pm0.040$ & $\cdots$ & $19.482\pm0.040$ \\
\hline
\multicolumn{5}{c}{$g_{CSP}$ band} \\
56633.22 & $-$15.28 & $18.062\pm0.011$ &    0.036 & $18.026\pm0.011$ \\ 
56636.26 & $-$12.24 & $16.622\pm0.009$ &    0.010 & $16.612\pm0.009$ \\ 
56637.11 & $-$11.39 & $16.330\pm0.006$ &    0.004 & $16.326\pm0.006$ \\ 
56639.11 &  $-$9.39 & $15.769\pm0.008$ & $-$0.006 & $15.775\pm0.008$ \\ 
56640.17 &  $-$8.33 & $15.555\pm0.007$ & $-$0.009 & $15.564\pm0.007$ \\ 
56643.15 &  $-$5.35 & $15.104\pm0.008$ & $-$0.014 & $15.118\pm0.008$ \\ 
56644.14 &  $-$4.36 & $15.011\pm0.007$ & $-$0.014 & $15.025\pm0.007$ \\ 
56645.13 &  $-$3.37 & $14.938\pm0.007$ & $-$0.013 & $14.951\pm0.007$ \\ 
56646.12 &  $-$2.38 & $14.890\pm0.009$ & $-$0.011 & $14.901\pm0.009$ \\ 
56647.16 &  $-$1.34 & $14.857\pm0.007$ & $-$0.009 & $14.866\pm0.007$ \\ 
56648.13 &  $-$0.37 & $14.819\pm0.007$ & $-$0.006 & $14.825\pm0.007$ \\ 
56649.11 &     0.61 & $14.826\pm0.006$ & $-$0.002 & $14.828\pm0.006$ \\ 
56650.15 &     1.65 & $14.830\pm0.007$ &    0.002 & $14.828\pm0.007$ \\ 
56651.10 &     2.60 & $14.850\pm0.008$ &    0.006 & $14.844\pm0.008$ \\ 
56652.10 &     3.60 & $14.895\pm0.009$ &    0.011 & $14.884\pm0.009$ \\ 
56653.13 &     4.63 & $14.925\pm0.005$ &    0.016 & $14.909\pm0.005$ \\ 
56654.18 &     5.68 & $14.982\pm0.006$ &    0.022 & $14.960\pm0.006$ \\ 
56655.19 &     6.69 & $15.040\pm0.009$ &    0.028 & $15.012\pm0.009$ \\ 
56656.20 &     7.70 & $15.159\pm0.009$ &    0.034 & $15.125\pm0.009$ \\ 
56657.21 &     8.71 & $15.202\pm0.010$ &    0.039 & $15.163\pm0.010$ \\ 
56658.14 &     9.64 & $15.266\pm0.008$ &    0.045 & $15.221\pm0.008$ \\ 
56659.15 &    10.65 & $15.357\pm0.006$ &    0.051 & $15.306\pm0.006$ \\ 
56660.11 &    11.61 & $15.448\pm0.008$ &    0.056 & $15.392\pm0.008$ \\ 
56661.16 &    12.66 & $15.534\pm0.007$ &    0.061 & $15.473\pm0.007$ \\ 
56662.12 &    13.62 & $15.663\pm0.007$ &    0.066 & $15.597\pm0.007$ \\ 
56663.11 &    14.61 & $15.741\pm0.007$ &    0.071 & $15.670\pm0.007$ \\ 
56664.10 &    15.60 & $15.870\pm0.010$ &    0.075 & $15.795\pm0.010$ \\ 
56665.08 &    16.58 & $15.955\pm0.008$ &    0.079 & $15.876\pm0.008$ \\ 
56666.06 &    17.56 & $16.071\pm0.011$ &    0.083 & $15.988\pm0.011$ \\ 
56667.07 &    18.57 & $16.154\pm0.010$ &    0.086 & $16.068\pm0.010$ \\ 
56669.06 &    20.56 & $16.359\pm0.013$ &    0.090 & $16.269\pm0.013$ \\ 
56670.14 &    21.64 & $16.491\pm0.013$ &    0.092 & $16.399\pm0.013$ \\ 
56671.06 &    22.56 & $16.562\pm0.011$ &    0.093 & $16.469\pm0.011$ \\ 
56672.12 &    23.62 & $16.662\pm0.009$ &    0.094 & $16.568\pm0.009$ \\ 
56673.09 &    24.59 & $16.739\pm0.009$ &    0.095 & $16.644\pm0.009$ \\ 
56675.05 &    26.55 & $16.893\pm0.008$ &    0.095 & $16.798\pm0.008$ \\ 
56676.11 &    27.61 & $16.970\pm0.007$ &    0.095 & $16.875\pm0.007$ \\ 
56677.04 &    28.54 & $17.033\pm0.009$ &    0.094 & $16.939\pm0.009$ \\ 
56678.11 &    29.61 & $17.089\pm0.009$ &    0.094 & $16.995\pm0.009$ \\ 
56679.05 &    30.55 & $17.166\pm0.006$ &    0.093 & $17.073\pm0.006$ \\ 
56680.12 &    31.62 & $17.233\pm0.008$ &    0.092 & $17.141\pm0.008$ \\ 
56681.05 &    32.55 & $17.293\pm0.008$ &    0.091 & $17.202\pm0.008$ \\ 
56684.09 &    35.59 & $17.430\pm0.008$ &    0.088 & $17.342\pm0.008$ \\ 
56685.07 &    36.57 & $17.484\pm0.008$ &    0.086 & $17.398\pm0.008$ \\ 
56686.05 &    37.55 & $17.492\pm0.010$ &    0.085 & $17.407\pm0.010$ \\ 
56687.06 &    38.56 & $17.511\pm0.009$ &    0.084 & $17.427\pm0.009$ \\ 
56690.08 &    41.58 & $17.617\pm0.010$ &    0.081 & $17.536\pm0.010$ \\ 
56692.08 &    43.58 & $17.674\pm0.011$ &    0.079 & $17.595\pm0.011$ \\ 
56694.06 &    45.56 & $17.696\pm0.009$ &    0.078 & $17.618\pm0.009$ \\ 
56696.07 &    47.57 & $17.733\pm0.012$ &    0.077 & $17.656\pm0.012$ \\ 
56703.07 &    54.57 & $17.906\pm0.019$ &    0.077 & $17.829\pm0.019$ \\ 
56708.06 &    59.56 & $17.968\pm0.011$ &    0.080 & $17.888\pm0.011$ \\ 
\hline
\multicolumn{5}{c}{$BV.5$ band}\\
56632.33 & $-$16.17 & $18.703\pm0.108$ &    0.090 & $18.613\pm0.108$ \\
56634.31 & $-$14.19 & $17.477\pm0.053$ &    0.109 & $17.367\pm0.053$ \\
56635.26 & $-$13.24 & $17.120\pm0.050$ &    0.116 & $17.003\pm0.050$ \\
56636.28 & $-$12.22 & $16.725\pm0.025$ &    0.122 & $16.602\pm0.025$ \\
56637.25 & $-$11.25 & $16.439\pm0.025$ &    0.126 & $16.312\pm0.025$ \\
56638.29 & $-$10.21 & $16.125\pm0.022$ &    0.130 & $15.995\pm0.022$ \\
56639.32 &  $-$9.18 & $15.868\pm0.022$ &    0.132 & $15.737\pm0.022$ \\
56640.29 &  $-$8.21 & $15.647\pm0.029$ &    0.133 & $15.514\pm0.029$ \\
56641.38 &  $-$7.12 & $15.426\pm0.025$ &    0.133 & $15.293\pm0.025$ \\
56645.36 &  $-$3.14 & $15.049\pm0.019$ &    0.130 & $14.920\pm0.019$ \\
56647.32 &  $-$1.18 & $14.994\pm0.014$ &    0.128 & $14.865\pm0.014$ \\
56648.31 &  $-$0.19 & $14.969\pm0.011$ &    0.128 & $14.841\pm0.011$ \\
56649.30 &     0.80 & $14.966\pm0.011$ &    0.128 & $14.839\pm0.011$ \\
56650.32 &     1.82 & $14.958\pm0.014$ &    0.128 & $14.830\pm0.014$ \\
56651.27 &     2.77 & $14.996\pm0.011$ &    0.128 & $14.868\pm0.011$ \\
56652.26 &     3.76 & $15.028\pm0.014$ &    0.127 & $14.901\pm0.014$ \\
56653.28 &     4.78 & $15.051\pm0.014$ &    0.127 & $14.924\pm0.014$ \\
56654.26 &     5.76 & $15.098\pm0.011$ &    0.126 & $14.972\pm0.011$ \\
56655.21 &     6.71 & $15.160\pm0.011$ &    0.124 & $15.036\pm0.011$ \\
56656.28 &     7.78 & $15.214\pm0.015$ &    0.122 & $15.093\pm0.015$ \\
56657.23 &     8.73 & $15.280\pm0.015$ &    0.118 & $15.162\pm0.015$ \\
56658.24 &     9.74 & $15.358\pm0.015$ &    0.114 & $15.244\pm0.015$ \\
56659.25 &    10.75 & $15.435\pm0.015$ &    0.109 & $15.326\pm0.015$ \\
56660.21 &    11.71 & $15.523\pm0.025$ &    0.104 & $15.419\pm0.025$ \\
56661.23 &    12.73 & $15.601\pm0.032$ &    0.097 & $15.504\pm0.032$ \\
56663.22 &    14.72 & $15.782\pm0.017$ &    0.081 & $15.701\pm0.017$ \\
56665.23 &    16.73 & $15.959\pm0.023$ &    0.063 & $15.896\pm0.023$ \\
56667.24 &    18.74 & $16.110\pm0.031$ &    0.032 & $16.078\pm0.031$ \\
56672.25 &    23.75 & $16.470\pm0.045$ & $-$0.110 & $16.580\pm0.045$ \\
56674.24 &    25.74 & $16.596\pm0.029$ & $-$0.136 & $16.732\pm0.029$ \\
56676.23 &    27.73 & $16.727\pm0.030$ & $-$0.151 & $16.878\pm0.030$ \\
56678.22 &    29.72 & $16.845\pm0.028$ & $-$0.161 & $17.006\pm0.028$ \\
56680.20 &    31.70 & $16.986\pm0.033$ & $-$0.166 & $17.152\pm0.033$ \\
56682.19 &    33.69 & $17.086\pm0.030$ & $-$0.168 & $17.254\pm0.030$ \\
56684.19 &    35.69 & $17.192\pm0.032$ & $-$0.167 & $17.358\pm0.032$ \\
56689.16 &    40.66 & $17.389\pm0.054$ & $-$0.153 & $17.542\pm0.054$ \\
56702.14 &    53.64 & $17.824\pm0.111$ & $-$0.085 & $17.909\pm0.111$ \\
\enddata
\tablenotetext{a}{Temporal phase given in days with respect to the time of $B$-band maximum light on MJD $= 56648.5\pm0.10$,  i.e., 2013 December 22.}
\end{deluxetable}`

\clearpage
\newcommand\p{\phantom{0}}
\begin{deluxetable}{cccccc}
\tabletypesize{\normalsize}
\tablewidth{0pt}
\tablecaption{Journal of Spectroscopic Observations.\label{specjor}}
\tablehead{
\colhead{Date} &
\colhead{MJD} &
\colhead{Phase$^{a}$} &
\colhead{Telescope$^{b}$} &
\colhead{Instrument} &
\colhead{$-v_{abs}$ $\ion{Si}{II} \lambda 6355$ [km s$^{-1}$]}}
\startdata
2013 Dec 07.5   & 56633.5        & $-15.0$       & FTS                           & FLOYDS$^{c}$               &  14,039$^{ +198}_{ -104}$    \\
2013 Dec 07.7   & 56633.7        & $-14.8$       & LJT                           & YFOSC                      &  13,709$^{ +274}_{ -151}$    \\
2013 Dec 07.8   & 56633.8        & $-14.7$       & Copernico                     & AFOSC                      &  13,732$^{ +510}_{ -132}$    \\
2013 Dec 08.5   & 56634.5        & $-14.0$       & FTS                           & FLOYDS$^{c}$               &  13,756$^{\p+90}_{\p-71}$    \\
2013 Dec 08.8   & 56634.8        & $-13.7$       & LJT                           & YFOSC                      &  13,322$^{ +326}_{ -142}$    \\
2013 Dec 09.7   & 56635.7        & $-12.8$       & LJT                           & YFOSC                      &  13,067$^{ +170}_{\p-61}$    \\
2013 Dec 14.0   & 56640.0        & $-8.5$        & Magellan Clay                 & MagE                       &  11,509$^{\p+76}_{\p-33}$    \\
2013 Dec 15.4   & 56641.4        & $-7.1$        & FTS                           & FLOYDS$^{c}$               &  11,174$^{ +146}_{\p-66}$    \\
2013 Dec 17.4   & 56643.4        & $-5.1$        & FTS                           & FLOYDS$^{c}$               &  10,645$^{\p+28}_{\p-28}$    \\
2013 Dec 20.5   & 56646.5        & $-2.0$        & FTS                           & FLOYDS$^{c}$               &  10,673$^{ +222}_{\p-24}$    \\
2013 Dec 20.7   & 56646.7        & $-1.8$        & LJT                           & YFOSC                      &  10,631$^{\p+52}_{\p-66}$    \\
2013 Dec 21.7   & 56647.7        & $-0.8$        & LJT                           & YFOSC                      &  10,178$^{\p+90}_{\p-47}$    \\
2013 Dec 25.8   & 56651.8        & $+3.3$        & NOT                           & ALFOSC                     &  10,149$^{\p+90}_{\p-71}$    \\
2013 Dec 27.7   & 56653.7        & $+5.2$        & XLT                           & BFOSC                      & \p9,862$^{ +415}_{\p-85}$    \\
2013 Dec 28.3   & 56654.3        & $+5.8$        & Keck                          & LRIS                       & \p9,913$^{\p+71}_{\p-14}$    \\
2013 Dec 31.2   & 56657.2        & $+8.7$        & Magellan Baade                & IMACS                      & \p9,734$^{\p+94}_{\p-57}$    \\
2013 Dec 31.7   & 56657.7        & $+9.2$        & LJT                           & YFOSC                      & \p9,682$^{\p+47}_{\p-57}$    \\
2014 Jan 03.6   & 56660.6        & $+12.1$       & XLT                           & BFOSC                      &  10,971$^{ +836}_{ -850}$    \\
2014 Jan 07.6   & 56664.6        & $+16.1$       & LJT                           & YFOSC                      & \p9,267$^{ +170}_{\p-94}$    \\
2014 Jan 08.3   & 56665.3        & $+16.8$       & du Pont                       & WFCCD                      & \p9,059$^{\p+99}_{\p-99}$    \\
2014 Jan 14.0   & 56671.0        & $+22.5$       & NOT                           & ALFOSC                     & \p8,705$^{ +302}_{ -123}$    \\
2014 Jan 23.3   & 56680.3        & $+31.8$       & Shane                         & Kast                       & \p8,134$^{\p+94}_{\p-33}$    \\
2014 Feb 09.9   & 56697.9        & $+49.4$       & NOT                           & ALFOSC                     & \p7,487$^{ +439}_{ -482}$    \\        
2014 Aug 14.5   & 56883.5        & $+$235.0        & Magellan Baade                & IMACS                      & $\cdots$                  \\
\enddata
\tablenotetext{a}{Temporal phase given in days with respect to the time of $B$-band maximum light on MJD $= 56648.5\pm0.10$, i.e., 2013 December 22.}
\tablenotetext{b}{FTS = 2.0-m Faulkes Telescope South, LJT = 2.4-m Li-Jiang Telescope,  Copernico = INAF Osservatorio Astronomico di Padova 1.8-m
Telescope, NOT = 2.5-m Nordic Optical Telescope, XLT= 2.6-m Xing-Long Telescope.}
\tablenotetext{c}{Spectra were published by \cite{graham17}.}
\end{deluxetable}

\clearpage
 \begin{deluxetable}{ccccccccc}
\tabletypesize{\tiny}
\tablewidth{0pt}
\tablecaption{Key parameters of SN~2013gy.\label{results}}
\tablehead{
\colhead{$t_{Bmax}$} &
\colhead{$\Delta$m$_{15}(B)$ } &
\colhead{$s_{BV}$} &
\colhead{$E(B-V)_{tot}$} &
\colhead{$\mu$$^a$} &
\colhead{$M_B$} &
\colhead{$t_{first}$}&
	\colhead{$\alpha$} &
	\colhead{$t_{rise}$} \\
\colhead{[MJD]} &
\colhead{[mag] } &
\colhead{} &
\colhead{[mag]} &
\colhead{[mag]} &
\colhead{[mag]} &
\colhead{[MJD]} &
\colhead{} &
\colhead{[days]} }
\startdata
56648.5$\pm0.10$   & $1.234\pm0.060$  & $0.892\pm0.05$ & $0.106\pm0.060$ & $33.75\pm0.15$  &  $-19.32\pm0.16$ & $56629.4\pm0.13$ &  $2.16^{+0.06}_{-0.06}$ & $19.10^{+0.12}_{-0.13}$ \\ 
\enddata
\tablenotetext{a}{Based on red-shift assuming $H_0 = 73$ km~s$^{-1}$ and adopting a correction based on a Virgo, Great Attractor and Shapley infall model.}
\end{deluxetable}

\clearpage

\begin{figure}
\setcounter{figure}{0}
  \includegraphics[width=0.8\textwidth]{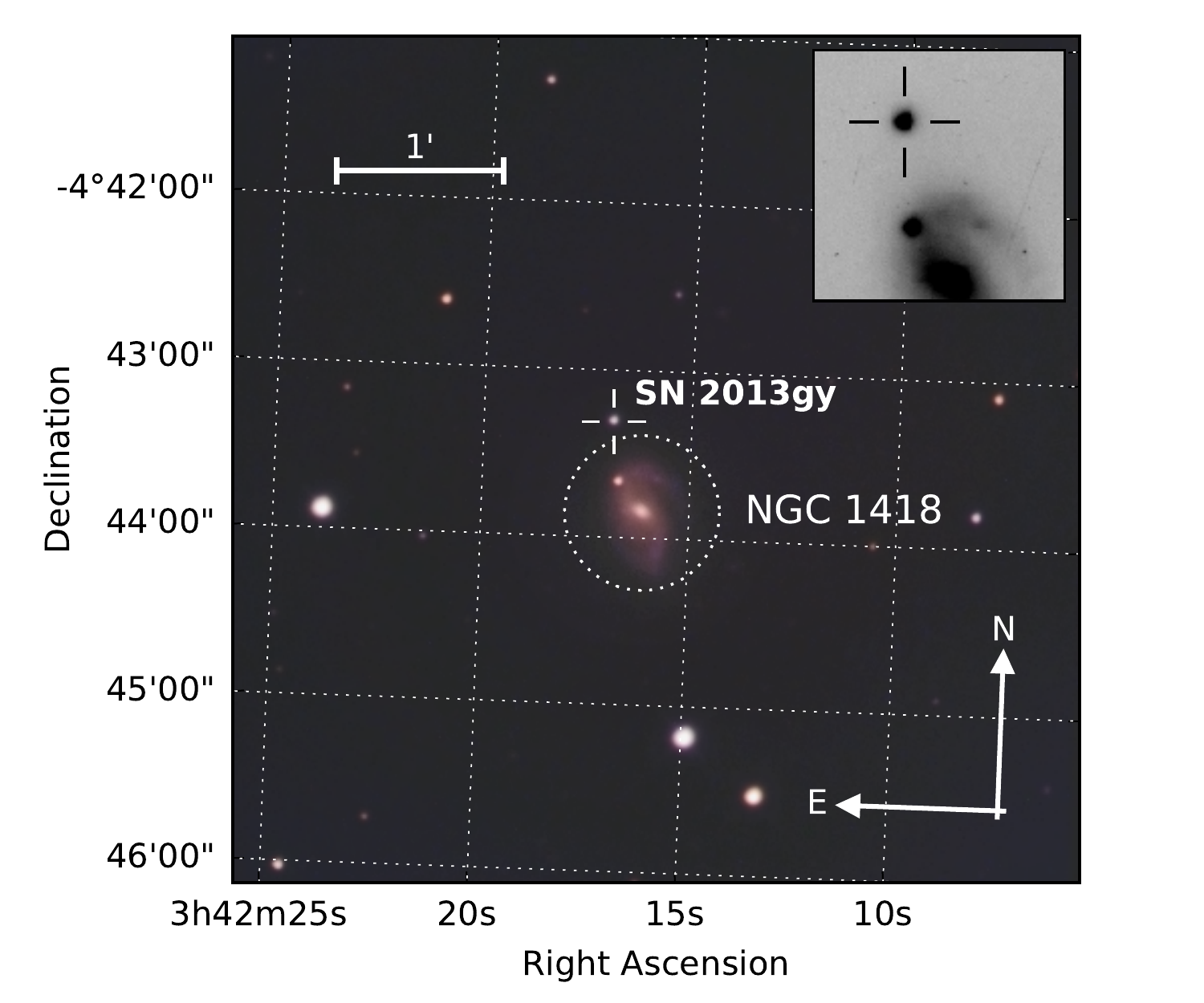}
  \caption{\textit{(online figure)} Composite image of NGC~1418  (North up and East left) constructed from multiband images obtained with the Swope telescope. 
    The inset contains a close-up view of the position of SN~2013gy, highlighting its position in the outer region of the host.}
  \label{fig:findingchart}
\end{figure}

\clearpage

\begin{figure}
\setcounter{figure}{1}
  \includegraphics[width=0.8\textwidth]{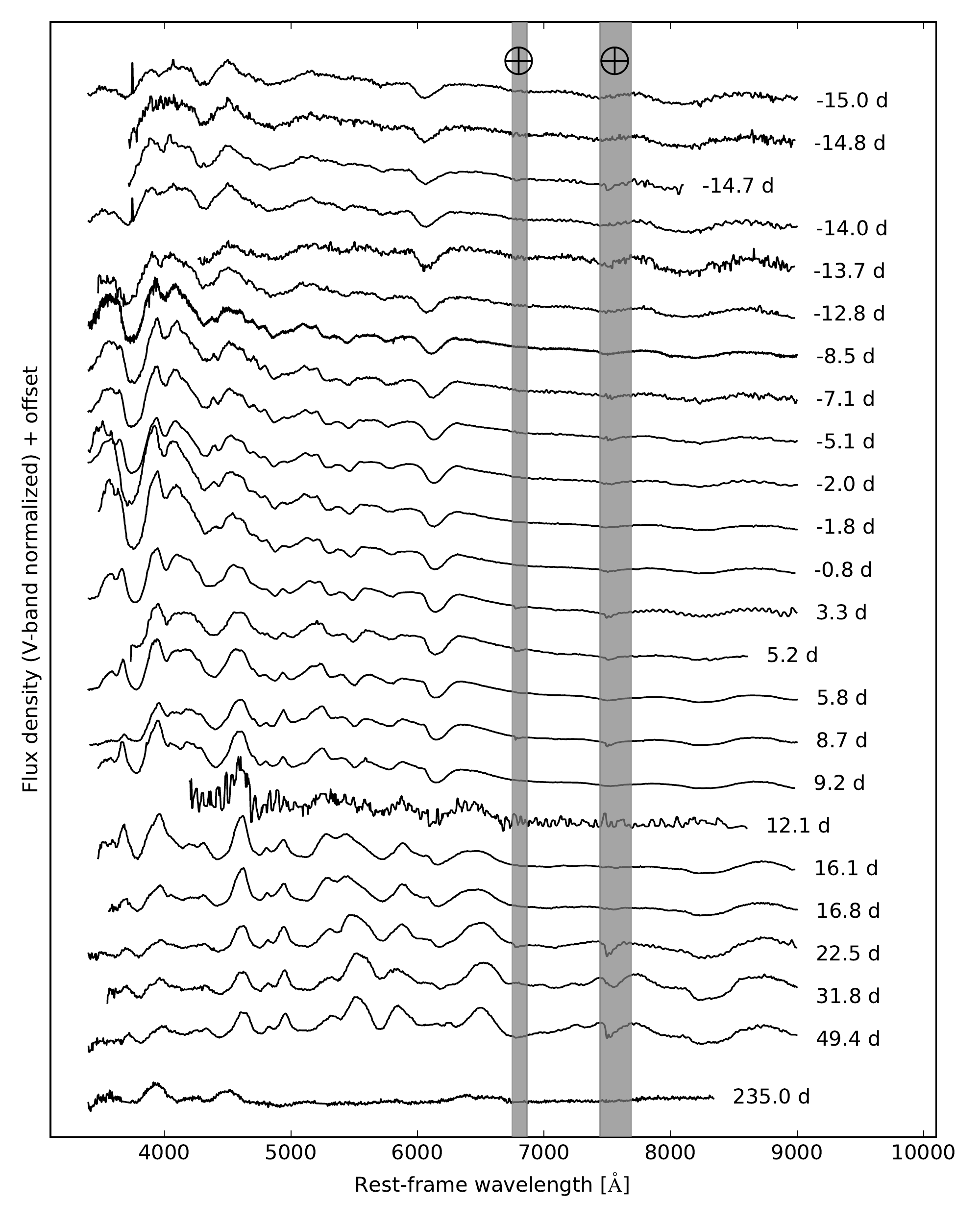}
  \caption{\textit{(online figure)} Spectroscopic time series of SN~2013gy used to compute S-corrections, enabling us to transform natural-system $g_{\rm CSP}$- and $BV.5$-band photometry to the $g_{\rm PS1}$ natural system.}
\label{fig:spectra}
\end{figure}

\end{document}